# Interfacial Resonance States-Induced Negative Tunneling Magneto-resistance in Orthogonally-Magnetized CoFeB/MgO/CoFeB

Puyang Huang,[1] Aitian Chen,[2] Jianting Dong,[3] Di Wu,[4] Xinqi Liu,[5] Zhenghang Zhi,[1] Jiuming Liu,[1] Albert Lee,[4] Bin Fang,[2] Jia Zhang,[3] Xi-Xiang Zhang,[2, a)] and Xufeng Kou [1,5, a)]

[1] *School of Information Science and Technology, ShanghaiTech University, Shanghai, 201210, China*

[2] *Physical Science and Engineering Division, King Abdullah University of Science and Technology, Thuwal, 23955-6900, Saudi Arabia*

[3] *School of Physics and Wuhan National High Magnetic Field Center, Huazhong University of Science and Technology, Wuhan, 430074, China*

[4] *Suzhou Inston Technology Co., Ltd., Suzhou, 215121, China.*

[5] *ShanghaiTech Laboratory for Topological Physics, ShanghaiTech University, Shanghai 201210, China.*

___________________________

[a)] Email: kouxf@shanghaitech.edu.cn, xixiang.zhang@kaust.edu.sa

**Abstract**

Magnetic tunneling junctions (MTJs) are essential for non-volatile magneto-resistive random access memory (MRAM) applications. Here, we report the observation of a large negative tunneling magneto-resistance (TMR) in the CoFeB/MgO/CoFeB system with an orthogonally-magnetized configuration. Through the thickness modulation of the MgO barrier, the negative TMR component can be enhanced up to 20% under a negative voltage bias. Moreover, the tunnel anisotropic magneto-resistance measurements unveil that the negative TMR component likely arises from the interfacial resonance states (IRS) in the minority band of the bottom ferromagnetic layer. Complementary first principle calculations further quantify the IRS location and strength with respect to the Fermi level position. Our work not only confirm the vital role of IRS in the electrical transport of MTJ, but also provide valuable insights for the design of new-generation voltage-controlled MRAM and related spintronic applications.

Magnetic tunnel junctions (MTJs), the key component of a magneto-resistive random access memory (MRAM) device, have been extensively investigated in the past four decades[1-5]. Given that the performance of MTJs relies heavily on their tunneling magneto-resistance (TMR)[6-10], previous work has identified the close correlation between the TMR ratio and the interfacial spin-polarized density of states (DOS) at the ferromagnetic (FM) layer/tunneling barrier hetero-interface[11-13]. In general, the interfacial DOS is associated with interfacial phenomena, which have both extrinsic and intrinsic origins[14, 15]. Because of the imperfect nature of the synthesized MTJ film stack, the relevant TMR ratio is strongly dependent on defects, interdiffusion, and interface roughness (*i.e.*, which also cause device-to-device variation, yet are difficult to be quantified)[1, 16]. In addition to extrinsic factors, interfacial resonance states (IRS) are believed to play a fundamental role in determining TMR[12, 17-20]. Specifically, when a large number of DOS appear in the minority band of the FM layer at a certain Fermi energy, the resultant IRS are formed, which in turn modify the transport characteristics of minority carriers. Consequently, once the contribution from the minority band surpasses the majority one, a negative TMR can be observed, as previously reported in epitaxial Fe/MgO/Fe MTJs[17] and NiFe/$Al_2O_3$/$Ta_2O_5$/NiFe systems[21]. In the meanwhile, the electrical properties of MTJ are also dependent on the tunneling barrier where evanescent states with complex wave vectors can interfere with electron transmission and decay of different Bloch states[12, 13]. Accordingly, it is proposed that the increase of the barrier thickness would enhance the IRS strength and the negative TMR effect[12, 22]. Given the emergence of voltage-controlled magnetic anisotropy (VCMA)-based MRAM, which typically requires a thick MgO tunneling barrier[23-25], it is therefore imperative to explore the IRS-induced magneto-transport responses in thick-barrier MTJs.

In this work, we study the thickness-dependent TMR in CoFeB/MgO($t$ nm)/CoFeB-based MTJs with orthogonally-magnetized configuration. Contrary to the control sample with a 2.5 nm-thick MgO barrier, we observe a distinctive magnetic field-induced upturned TMR slope in the $t$ = 3 nm sample when subjected to a large negative gate voltage, indicating the presence of a negative TMR component.

Moreover, the voltage-dependent tunnel anisotropic magneto-resistance (TAMR) mapping of the CoFeB/MgO(3 nm)/CoFeB sample exhibits a notable twofold-to-fourfold symmetry transition with the same onset voltage bias, hence manifesting the impact of IRS at the bottom MgO/CoFeB interface on the DOS coupling. Additionally, the *ab initio* electronic structure calculations further quantify the interplay between the MgO layer thickness and the IRS strength in the minority band. Our findings underscore the significance of IRS in shaping the TMR ratio, and may provide more insights into designing suitable MTJ structure for high-performance VCMA-MRAM.

The orthogonally-magnetized MTJs were grown on a thermally oxidized Si/SiO$_2$ substrate using a Singulus ROTARIS magnetron sputtering system. The MTJs featured a film stack comprising Ru(5 nm)/Ta(7 nm)/Co$_{40}$Fe$_{40}$B$_{20}$(2.5 nm)/MgO($t$ nm)/Co$_{40}$Fe$_{40}$B$_{20}$(1 nm)/Ta(5 nm), as shown in Fig. 1(a). The 1 nm-thick bottom CoFeB layer serves as the free layer (*i.e.*, with perpendicular magnetic anisotropy (PMA) $M_1$), while the 2.5 nm-thick top CoFeB layer acts as the fixed layer (*i.e.*, with in-plane magnetic anisotropy (IMA) $M_2$). To ensure the high quality of the tunneling barrier and stable magnetizations of the FM layers, the CoFeB/MgO/CoFeB film underwent a post-annealing process in vacuum at 250 °C for 30 min[26]. The cross-sectional high-resolution transmission electron microscope (HRTEM) image of the MTJ reveals both a well-ordered crystalline structure of the MgO (001) barrier layer and clear CoFeB/MgO interfaces. In addition, the colored EDX mappings in Fig. 1(b) demonstrate the uniform distribution of each constituent element without detectable atom intermixing/diffusion.

Following the growth and structural characterization of the sample, the film was patterned into 18 μm × 6 μm elliptical MTJ devices by using the standard nano-fabrication process. Afterwards, magneto-transport measurements were conducted using the two-probe method. As illustrated in Fig. 1(c), the gate voltage $V_g$ was applied from the top to the bottom electrode (*i.e.*, given the capacitive nature of the thick MgO barrier used in this work, the applied gate voltage is able to tune the Fermi level at the MgO/CoFeB interface, and the $V_g$-dependent TMR is concurrently measured between the two electrodes), while an external in-plane magnetic field ($B$) was utilized to modulate the magnetization of the bottom $Co_{40}Fe_{40}B_{20}$(1 nm) layer. According to the Julliere equation, the conductance of the MTJ is expressed as $G = G_{surface}(1 + P_{f1}P_{f2}cos\theta)$, where $G_{surface}$ is the mean surface conductance, $P_{f1}$ and $P_{f2}$ are the effective polarizations of the two CoFeB layers, and $\theta$ is the angle between $M_1$ and $M_2$[27]. Experimentally, the tunneling resistance ($R=1/G$) of the $t = 2.5$ nm device monotonically decreases with the applied magnetic field (Fig. 2(a)), reaching its minimum value ($R_{min}$) when $B > 250$ mT (i.e., above this critical field $B_{min}$, $M_1$ is parallel to $M_2$, and $\theta = 0°$). In the meantime, the voltage-dependent TMR curves display a typical VCMA behavior, namely a smaller $B_{min}$ is able to drive the two-terminal tunneling resistance into the $R_{min}$ state under a large positive $V_g$[28-30]. On the contrary, the TMR slopes of the $t = 3$ nm MTJ device exhibit a distinct evolution trend concerning the applied magnetic field and bias voltage. As highlighted in Fig. 2(b), when a large negative voltage is applied (i.e., $V_g < -600$ mV), the measured tunneling resistance no longer remains at the $R_{min}$ state under high magnetic fields; instead, it gradually increases with $B$ even the magnetic moments of two FM layers are already parallel to each other. Additionally, the characteristic field $B_{min}$ with respect to $R_{min}$ reduces from 317.5 mT (−800 mV) to 278.9 mT (−1000 mV) (inset of Fig. 2(b)), which is opposite to the $t = 2.5$ nm counterpart. Quantitatively, the change of the two-terminal tunneling resistance is described by $\Delta R = (R_P - R_{min})/(R_0 - R_{min})$, where $R_P$ is the resistance at $B = 3000$ mT (i.e., $M_1 // M_2$), and $R_0$ is the zero-field resistance of the MTJ sample. As displayed in Fig. 2(c), the $\Delta R$ ratio increases under larger negative voltage biases until it reaches up to 20% at $V_g = -1000$ mV.

Considering the potential influence of IRS in thick MgO layers on the occurrence of negative TMR, we conducted a decoupling procedure to analyze the measured TMR curves of the $t = 3$ nm MTJ sample. The fitting result, exemplified in the inset of Fig. 2(d), demonstrates that the $R$-$B$ line-shape at $V = -900$ mV can be represented by a combination of positive and negative TMR slopes. Strikingly, the extracted pseudo-anisotropy fields $H_{k'}$ (i.e., defined as the signature magnetic field corresponding to the $\theta = 45°$ magnetization configuration of the two FM layers) from these two contributions exhibit opposite dependence on the applied $V_g$: the $H_{k'}$ of the negative (positive) TMR component contracts (expands) under the larger negative voltage, consistent with the positively correlated $B_{min}$-$V$ trend in the inset of Fig. 2(b). In fact, in our as-grown MTJ structure, the applied negative voltage facilitates the electron tunneling across the thick MgO barrier and results in the occupation of IRS at the bottom MgO/CoFeB interface[17, 19], as illustrated in the inset of Fig. 2(c). Consequently, the strengthened transport of minority carriers affects the magnetization of the bottom free CoFeB layer, leading to a more pronounced negative TMR component. It is worth noting that negative TMR effect may also be associated with spin-valley polarization[31, 32], or bulk/interface DOS coupling[17]. However, in the CoFeB layer, the valley polarization induced by the combination of inversion asymmetry and strong spin-orbit coupling (SOC) can be disregarded[33]. On the other hand, the bulk-interface DOS coupling normally requires a thick FM layer (~50 nm)[17], which is not applicable to our system neither.

To further validate the underlying TMR physics in our MTJ structure, voltage-dependent tunnel anisotropic magneto-resistance measurements were carried out in the physical property measurement system (PPMS)[34, 35]. Figure 3 summarizes the corresponding TAMR mappings of the two CoFeB/MgO($t$)/CoFeB samples with $t$ = 2.5 nm and 3 nm, respectively. By successively rotating the external magnetic field $|B|$ = 12 T (*i.e.*, which is large enough to align $M_1$ and $M_2$ along the same direction) from $\theta_B$ = 0 ° to 360 ° (*i.e.*, where $\theta_B$ is the angle between the magnetic field $B$ and the normal direction $\vec{n}$ of the film stack, as shown in Fig. 3(a)), the $V_g$-dependent TAMR results always exhibit a twofold symmetry in the $t$ = 2.5 nm device. Specifically, the TMR values reach their highest (lowest) states when the magnetizations of the two FM layers are along the perpendicular (in-plane) directions (Fig. 3(b)). In contrast, the TAMR polar plots of the $t$ = 3 nm sample retain the "horizontal 8"-shape contours under small gate voltages, until the twofold-to-fourfold symmetry transition occurs when $V_g \leq -600$ mV, as highlighted in Fig. 3(c). Based on the tight-binding model, once the majority-spin band $\Delta_1$ is coupled with a minority IRS via interfacial

SOC, the tunneling magneto-resistance of the MTJ will be modulated by the anisotropy of interface DOS[34]. In this scenario, the enhanced IRS strength, enabled by the large negative bias, may introduce additional tunneling states (*i.e.*, which are isotropic versus $\theta_B$) to modify the interfacial Green's function of IRS. Accordingly, this modification leads to the appearance of a fourfold petal-shape TAMR in the MTJ system with a thick MgO barrier[17]. More importantly, it is found that the onset voltage of $V_g = -600$ mV triggers both the twofold-to-fourfold TAMR transition (Fig. 3(c)) and the appearance of the negative TMR component (Fig. 2(b)) in the $t = 3$ nm sample, providing further evidence for the IRS-related mechanism underlying these two effects. Furthermore, we need to point out that the intrinsic anisotropic magneto-resistance behavior of the free/fixed CoFeB layers can be negligible because of the small SOC of CoFeB in the CoFeB/MgO/CoFeB structure[36, 37], hence ensuring that the measured resistance change is mainly from the TAMR response.

In order to evaluate the IRS strength with varied MgO thicknesses, the spin-polarized first-principles calculations were conducted by using the Vienna ab initio simulation package (VASP)[38]. In particular, the interfacial formation energy in a FeCo/MgO/FeCo supercell structure is computed by using the equation[39]:

$$\Delta E_{\text{CoFe/O}} = \frac{1}{2}(E_{sc} - 3E_{\text{CoFe}} - 2E_{\text{MgO}} - E_{\text{Co}} - E_{\text{Fe}} - 2E_{\text{O}} - 2E_{\text{Mg}})$$

where $E_{sc}$ is the total energy of the supercell, $E_{\text{CoFe}}$, $E_{\text{MgO}}$, $E_{\text{Co}}$, $E_{\text{Fe}}$, and $E_{\text{Mg}}$ correspond to the energies of each unit cell of CoFe, MgO, Co, Fe, and Mg, and $E_{\text{O}}$ represents the energy of half an oxygen molecule. Figure 4(a) depicts the most stable interface configuration from simulation where the spacing between the Co and Fe atoms is 2.15 Å, and Fe is directly opposite to O at the hetero-interface, with an interfacial formation energy of 7.93 eV. Subsequently, the projected density of states (PDOS) were calculated using a cutoff energy of 600 eV and the $10 \times 10 \times 1$ and $20 \times 20 \times 1$ *k*-point meshes in the Brillouin zone. To match the as-grown MTJ film stack, the MgO barrier thicknesses in the FeCo/MgO/FeCo supercells were chosen as 11 layers (2.22 nm), 13 layers (2.66 nm) and 15 layers (3.11 nm). In addition, the in-plane lattice constant of these supercells was fixed at $a = 4.02$ Å, while the total supercell lengths were 37.88 Å, 42.32 Å and 46.76 Å, respectively.

Previous studies have suggested that the majority channel conductance of the MTJ structure is mostly determined by Bloch states with the $\Delta_1$ ($s$, $p_z$, $d_z^2$) symmetry; while the minority channel conductance is primarily influenced by $\Delta_5$ ($p_x$, $p_y$, $d_{xz}$, $d_{yz}$) states at both CoFe/MgO interfaces[12]. Accordingly, Fig. 4(b) presents the calculated PDOS of both majority and minority bands (*i.e.*, $\Delta_1$ and $\Delta_5$ symmetries) for the three FeCo/MgO($t$)/FeCo supercells. It is seen that two IRS peaks, S1 and S2 (*i.e.*, both of which predominantly belong to the $\Delta_5$ symmetry), are developed in the minority bands of all PDOS spectra, and their locations are 0.2 eV and 0.8 eV above the Fermi level (*i.e.*, which are consistent with previous reports[19]). In accordance with the physical model depicted in Fig. 2(c), the tunneling electrons occupying the IRS contribute to the minority carrier transport as well as change the anisotropy of the interface DOS. As the overall S2 amplitude ($\Delta_1$ + $\Delta_5$) increases with the MgO thickness (Fig. 4(b)), it implies that more available IRS at the bottom MgO/CoFeB interface can capture electrons under large negative gate voltages (*i.e.*, the Fermi level will be tuned towards S1 and S2 when $V_g < 0$ V), therefore leading to a pronounced negative TMR effect and a fourfold TAMR pattern in the $t = 3$ nm MTJ device.

In conclusion, our investigation delved into the unconventional TMR effect and its underlying mechanism in the CoFeB/MgO/CoFeB-based MTJ structure with varied MgO thicknesses and gate biases. The observations of the high magnetic field-induced upturned TMR and the twofold-to-fourfold TAMR symmetry transition under negative $V_g$ can be attributed to the enhanced IRS at the bottom MgO/CoFeB interface. Meanwhile, the consistency between our theoretical calculations and experimental findings highlights the importance of considering IRS when studying MTJs with a thick MgO barrier. Furthermore, our work showcases the complementary nature of TMR and TAMR as a reliable approach for IRS characterization, thus offering useful guidance for MTJ-based device design.

**Acknowledgments**

This work is sponsored by the National Key R&D Program of China (grant no. 2021YFA0715503), National Natural Science Foundation of China (grant no. 92164104), and the Major Project of Shanghai Municipal Science and Technology (grant no. 2018SHZDZX02), the Shanghai Engineering Research Center of Energy Efficient and Custom AI IC, and the ShanghaiTech Quantum Device and Soft Matter Nano-fabrication Labs (SMN180827). The work at King Abdullah University of Science and Technology (KAUST) was supported by KAUST Office of Sponsored Research (OSR) under Award Nos. ORA-CRG8-2019-4081 and ORA-CRG10-2021-4665. X.K. acknowledges support from the Shanghai Rising-Star Program (grant no. 21QA1406000) and the Open Fund of State Key Laboratory of Infrared Physics

**Data Availability**

AIP Publishing believes that all datasets underlying the conclusions of the paper should be available to readers. We encourage authors to deposit their datasets in publicly available repositories (where available and appropriate) or present them in the main manuscript. All research articles must include a data availability statement informing where the data can be found. By data we mean the minimal dataset that would be necessary to interpret, replicate and build upon the findings reported in the article. The data that support the findings of this study are available from the corresponding author upon reasonable request.

**References:**

[1] H. Yang, S. O. Valenzuela, M. Chshiev, S. Couet, B. Dieny, B. Dlubak, A. Fert, K. Garello, M. Jamet, D.-E. Jeong, K. Lee, T. Lee, M.-B. Martin, G. S. Kar, P. Sénéor, H.-J. Shin and S. Roche, Nature **606** (7915), 663 (2022).

[2] B. Dieny, I. L. Prejbeanu, K. Garello, P. Gambardella, P. Freitas, R. Lehndorff, W. Raberg, U. Ebels, S. O. Demokritov, J. Akerman, A. Deac, P. Pirro, C. Adelmann, A. Anane, A. V. Chumak, A. Hirohata, S. Mangin, S. O. Valenzuela, M. C. Onbaşlı, M. d'Aquino, G. Prenat, G. Finocchio, L. Lopez-Diaz, R. Chantrell, O. Chubykalo-Fesenko and P. Bortolotti, Nature Electronics **3** (8), 446 (2020).

[3] X. Dong, C. Xu, Y. Xie and N. P. Jouppi, IEEE Transactions on Computer-Aided Design of Integrated Circuits and Systems **31** (7), 994 (2012).

[4] S. Ikeda, J. Hayakawa, Y. M. Lee, F. Matsukura, Y. Ohno, T. Hanyu and H. Ohno, IEEE Transactions on Electron Devices **54** (5), 991 (2007).

[5] D. Apalkov, B. Dieny and J. M. Slaughter, Proceedings of the IEEE **104** (10), 1796 (2016).

[6] M. Julliere, Phys. Rev. A **54** (3), 225 (1975).

[7] S. S. P. Parkin, C. Kaiser, A. Panchula, P. M. Rice, B. Hughes, M. Samant and S.-H. Yang, Nat. Mater. **3** (12), 862 (2004).

[8] S. Yuasa, T. Nagahama, A. Fukushima, Y. Suzuki and K. Ando, Nat. Mater. **3** (12), 868 (2004).

[9] S. E. Lee, J. U. Baek and J. G. Park, Sci. Rep. **7** (1), 11907 (2017).

[10] J. G. Park, T. H. Shim, K. S. Chae, D. Y. Lee, Y. Takemura, S. E. Lee, M. S. Jeon, J. U. Baek, S. O. Park and J. P. Hong, IEEE IEDM., (2014).

[11] M. Ansarino, H. M. Moghaddam and A. Bahari, Physica E (2019).

[12] W. H. Butler, X. G. Zhang, T. C. Schulthess and J. M. Maclaren, Phys. Rev. B: Condens. Matter **63** (5) (2001).

[13] H. X. Yang, M. Chshiev, B. Dieny, J. H. Lee, A. Manchon and K. H. Shin, Phys. Rev. **84** (5), p.054401.1 (2011).

[14] A. Soumyanarayanan, N. Reyren, A. Fert and C. Panagopoulos, Nature **539** (7630), 509 (2016).

[15] F. Hellman, A. Hoffmann, Y. Tserkovnyak, G. S. D. Beach, E. E. Fullerton, C. Leighton, A. H. MacDonald, D. C. Ralph, D. A. Arena, H. A. Dürr, P. Fischer, J. Grollier, J. P. Heremans, T. Jungwirth, A. V. Kimel, B. Koopmans, I. N. Krivorotov, S. J. May, A. K. Petford-Long, J. M. Rondinelli, N. Samarth, I. K. Schuller, A. N. Slavin, M. D. Stiles, O. Tchernyshyov, A. Thiaville and B. L. Zink, Reviews of Modern Physics **89** (2), 025006 (2017).

[16] Y. Ohsawa, N. Shimomura, T. Daibou, Y. Kamiguchi, S. Shirotori, T. Inokuchi, D. Saida, B. Altansargai, Y. Kato, H. Yoda, T. Ohkubo and K. Hono, IEEE Transactions on Magnetics **52** (7), 1 (2016).

[17] C. Tiusan, J. Faurevincent, C. Bellouard, M. Hehn, E. Jouguelet and A. Schuhl, Phys. Rev. Lett. **93** (10), 106602 (2004).

[18] Y. Lu, H. X. Yang, C. Tiusan, M. Hehn, M. Chshiev, A. Duluard, B. Kierren, G. Lengaigne, D. Lacour and C. Bellouard, Phys. Rev. B **86** (18), 184420 (2012).

[19] P. J. Zermatten, G. Gaudin, G. Maris, M. Miron and M. Hehn, Phys. Rev. B **78** (3), 908 (2008).

[20] I. Rungger, O. Mryasov and S. Sanvito, Phys. Rev. B **79** (9), 094414 (2009).

[21] M. Sharma, S. X. Wang and J. H. Nickel, Phys. Rev. Lett. **82** (3), 616 (1999).

[22] K. D. Belashchenko, J. Velev and E. Y. Tsymbal, Phys. Rev. B **72** (14), 404 (2005).

[23] E. Grimaldi, V. Krizakova, G. Sala, F. Yasin, S. Couet, G. Sankar Kar, K. Garello and P. Gambardella, Nature Nanotechnology **15** (2), 111 (2020).

[24] D. Bhattacharya, S. A. Razavi, H. Wu, B. Dai, K. L. Wang and J. Atulasimha, Nature Electronics **3** (9), 539 (2020).

[25] S.-h. C. Baek, K.-W. Park, D.-S. Kil, Y. Jang, J. Park, K.-J. Lee and B.-G. Park, Nature Electronics **1** (7), 398 (2018).

[26]B. Cui, A. Chen, X. Zhang, B. Fang, Z. Zeng, P. Zhang, J. Zhang, W. He, G. Yu, P. Yan, X. Han, K. L. Wang, X. Zhang and H. Wu, Advanced Materials **n/a** (n/a), 2302350 (2023).
[27]J. C. Slonczewski, Phys. Rev. B **39** (10), 6995 (1989).
[28]Y. Shiota, F. Bonell, S. Miwa, N. Mizuochi, T. Shinjo and Y. Suzuki, Appl. Phys. Lett. **103** (8), 944 (2013).
[29]W.-G. Wang, M. Li, S. Hageman and C. L. Chien, Nat. Mater. **11** (1), 64 (2012).
[30]J. Zhu, J. A. Katine, G. E. Rowlands, Y.-J. Chen, Z. Duan, J. G. Alzate, P. Upadhyaya, J. Langer, P. K. Amiri, K. L. Wang and I. N. Krivorotov, Phys. Rev. Lett. **108** (19), 197203 (2012).
[31]C. Zhao, J. Ou, Z. Wen and W. Lu, Phys. Rev. A **452**, 128443 (2022).
[32]X. Tian, H. Wang, X. Qiu, Z. Cao, J. Hou and C. Yang, Physica E **142**, 115301 (2022).
[33]Z. Zhou, P. Marcon, X. Devaux, P. Pigeat, A. Bouché, S. Migot, A. Jaafar, R. Arras, M. Vergnat, L. Ren, H. Tornatzky, C. Robert, X. Marie, J.-M. George, H.-Y. Jaffrès, M. Stoffel, H. Rinnert, Z. Wei, P. Renucci, L. Calmels and Y. Lu, ACS Applied Materials & Interfaces **13** (27), 32579 (2021).
[34]L. Gao, X. Jiang, S. H. Yang, J. D. Burton, E. Y. Tsymbal and S. Parkin, Phys. Rev. Lett. **99** (22), 226602 (2007).
[35]A. N. Chantis, K. D. Belashchenko, E. Y. Tsymbal and M. van Schilfgaarde, Phys. Rev. Lett. **98** (4), 046601 (2007).
[36]J. Moser, A. Matos-Abiague, D. Schuh, W. Wegscheider, J. Fabian and D. Weiss, Phys. Rev. Lett. **99** (5), 056601 (2007).
[37]Y. A. Bychkov and E. I. Rashba, J. Phys. C: Solid State Phys. **17** (33), 6039 (1984).
[38]G. Kresse and J. Hafner, Phys. Rev. B **47** (1), 558 (1993).
[39]D. F. Shao, G. Gurung, T. R. Paudel and E. Y. Tsymbal, Phys. Rev. Mater. **3** (2) (2019).

**Figure Caption:**

FIG. 1. (a) Left panel: schematic of the orthogonally-magnetized MTJ structure which consists of Ru(5 nm)/Ta(7 nm)/Co$_{40}$Fe$_{40}$B$_{20}$(2.5 nm)/MgO($t$ nm)/Co$_{40}$Fe$_{40}$B$_{20}$(1 nm)/Ta(5 nm). Right panel: the cross-sectional HRTEM image of the MTJ film with sharp MgO/CoFeB interfaces. (b) EDX mapping visualizing the uniform distribution of the Co, Fe, and Mg elements, respectively. (c) Optical microscopy image of the top-gated MTJ device.

FIG. 2. Gate-dependent normalized tunneling magneto-resistance curves of the CoFeB/MgO($t$)/CoFeB-based MTJ device with (a) $t = 2.5$ nm and (b) $t = 3$ nm. Inset of Fig. 2(b): The change of $B_{min}$ with $-1000$ mV $\leq V_g \leq -800$ mV. (c) Gate-dependent high-field resistance change ratio $\Delta R$ of the $t = 3$ nm MTJ device. Inset: illustration of the electron tunneling under the negative bias. (d) Fitted pseudo-anisotropy field $H_{k'}$ of both negative and positive TMR components of the $t = 3$ nm sample. Inset: fitting result of the $V_g = -900$mV case.

FIG. 3. (a) Schematic of the TAMR measurement. The applied magnetic field of $B = 12$ T is used to align the fixed and free FM layers. (b)-(c) Gate-dependent TAMR polar plots of the (b) $t = 2.5$ nm and (c) $t = 3$ nm MTJ devices. The twofold-to-fourfold symmetry transition occurs in the CoFeB(2.5 nm)/MgO(3 nm)/CoFeB(1 nm) sample when $V_g \leq -600$ mV.

FIG. 4. (a) CoFe/MgO($t$)/CoFe tri-layer supercell used in the VASP calculations. (b) Comparisons of majority and minority PDOS with $\Delta_1$ and $\Delta_5$ symmetry of the supercell with $t = 2.2$ nm, 2.665 nm, 3.1 nm, where S1 and S2 represent two IRS at the minority bands. With the increase of the MgO thickness, the overall S2 amplitude ($\Delta_1 + \Delta_5$) also increases, indicating there are more IRS available to capture the tunneling electrons at the bottom MgO/CoFeB interface.

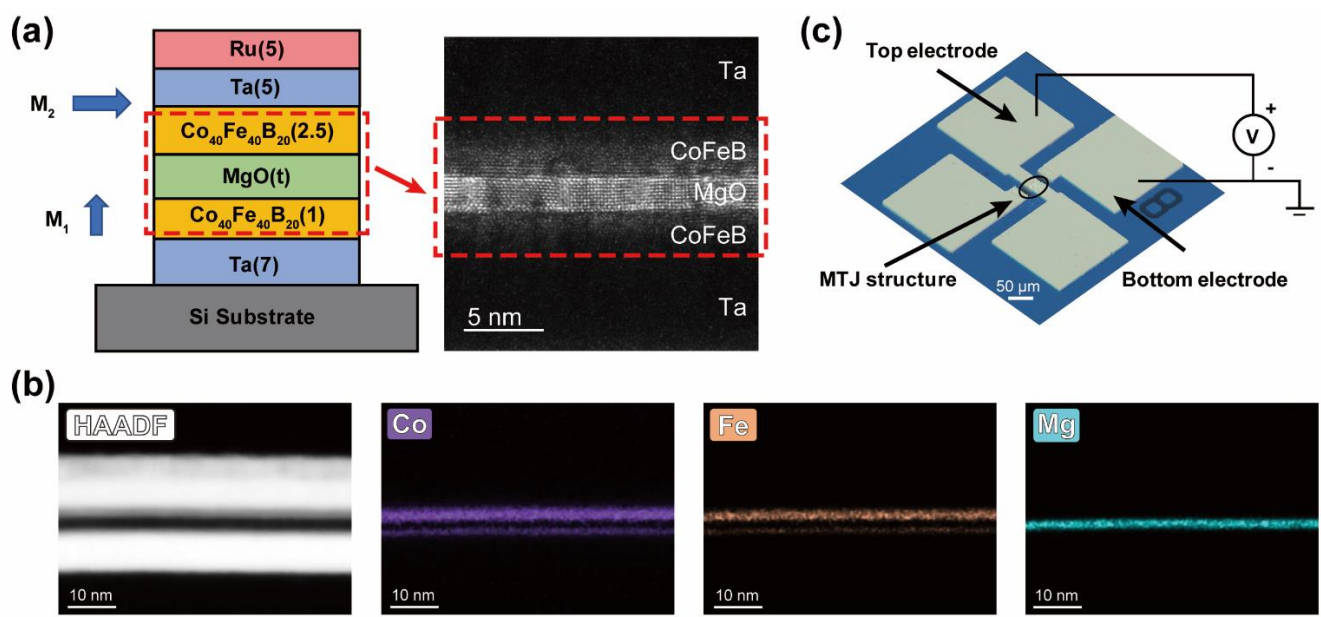

FIG. 1. Huang *et al.*

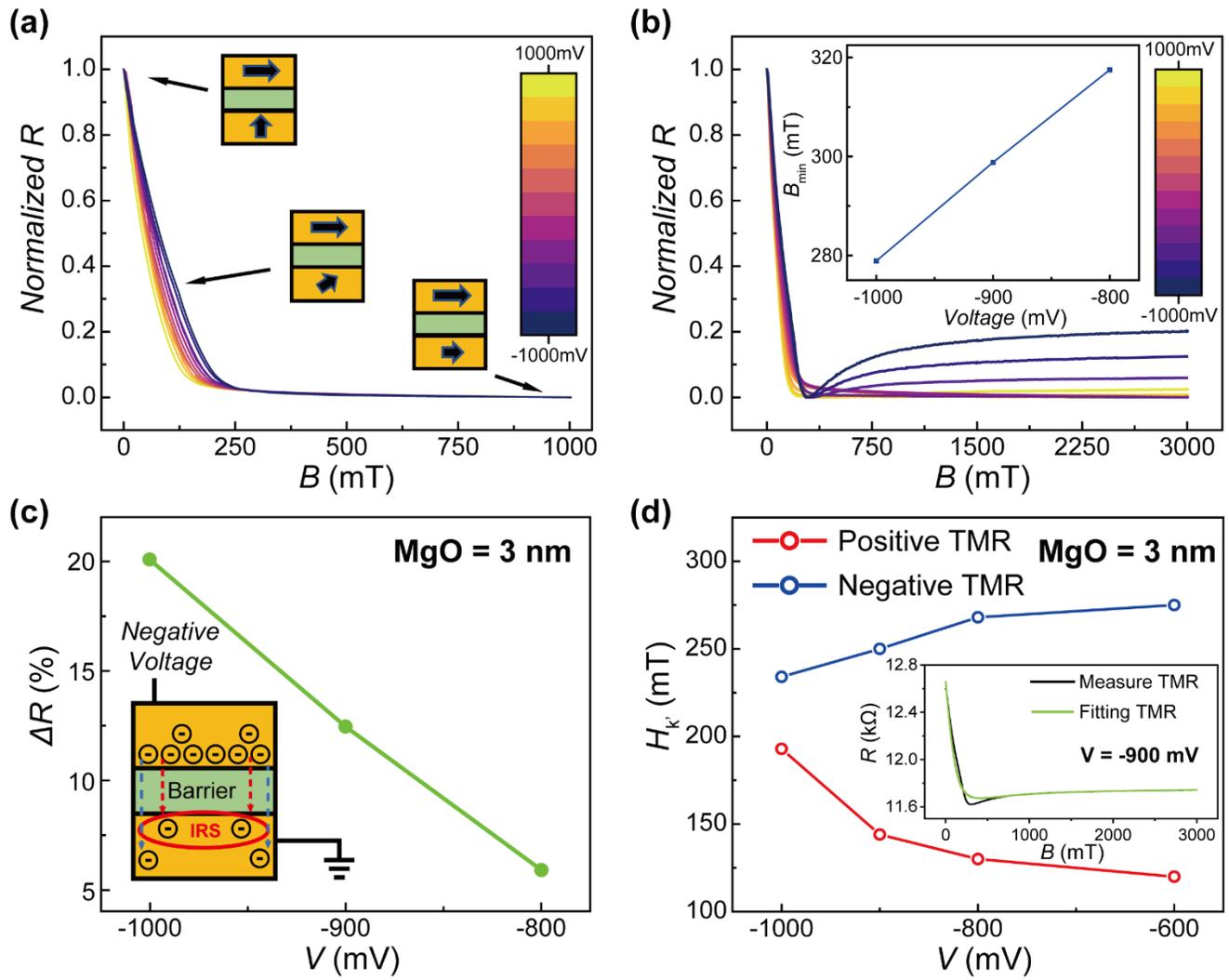

FIG. 2. Huang *et al.*

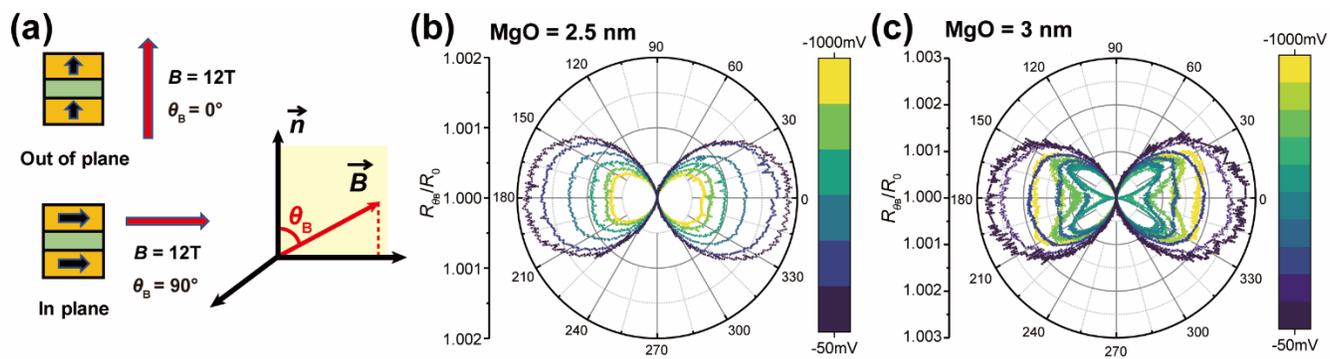

FIG. 3. Huang *et al.*

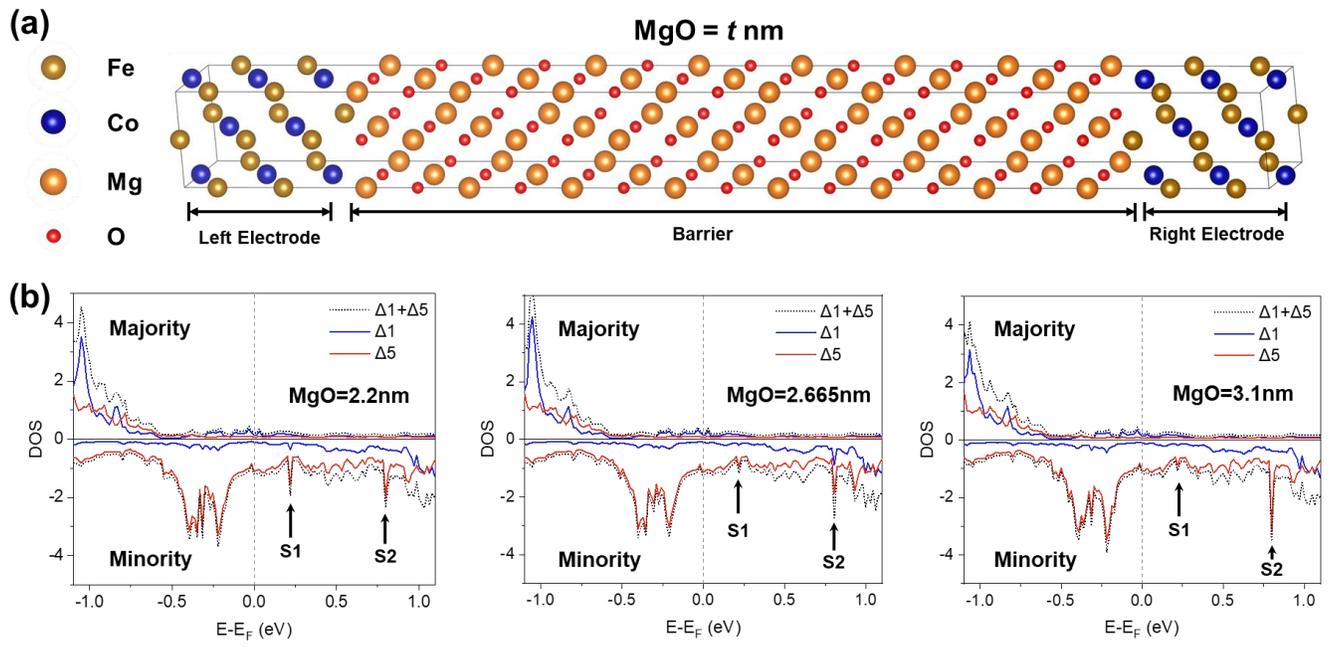

FIG. 4. Huang *et al.*